\documentclass[11pt,authoryear]{elsarticle}
\usepackage[utf8]{inputenc} 
\usepackage{xcolor}
\usepackage{soul} 

\usepackage{geometry} 
\usepackage{graphicx} 
\graphicspath{ {images/} }  

\usepackage{booktabs} 
\usepackage{array} 
\usepackage{paralist} 
\usepackage{verbatim} 
\usepackage{subfig} 

\usepackage{fancyhdr} 
\usepackage{sectsty}
\allsectionsfont{\sffamily\mdseries\upshape} 

\usepackage[nottoc,notlof,notlot]{tocbibind} 
\usepackage[titles,subfigure]{tocloft} 

\usepackage{etoolbox}
\apptocmd{\sloppy}{\hbadness 10000\relax}{}{}
\usepackage{bm} 
\usepackage{url} 
\usepackage{amsmath} 
\usepackage{amssymb}
\usepackage{multicol}
\usepackage{multirow}
\newcommand{\code}[1]{\texttt{#1}}

\begin{document}


\begin{frontmatter}

\title{Comparison of Gaussian process modeling software}

\author[NU]{Collin~B.~Erickson\corref{cor1}}
\ead{collinerickson@u.northwestern.edu}
\author[NU]{Bruce~E.~Ankenman} 
\ead{ankenman@northwestern.edu}
\author[NPS]{Susan~M.~Sanchez}
\ead{ssanchez@nps.edu}

\cortext[cor1]{Corresponding author}
\address[NU]{Department of Industrial Engineering and Management Sciences, Northwestern University, 2145~Sheridan~Rd., Evanston, IL, 60208 USA}
\address[NPS]{Department of Operations Research,
Naval Postgraduate School, 1411~Cunningham~Rd., Monterey,~CA, 93943 USA}

\begin{abstract}

Gaussian process fitting, or kriging, is often used to create a model from a set of data. Many available software packages do this, but we show that very different results can be obtained from different packages even when using the same data and model.
We describe the parameterization, features, and optimization used by eight different fitting packages that run on four different platforms.
We then compare these eight packages using various data functions and data sets, revealing that there are stark differences between the packages.
In addition to comparing the prediction accuracy, the predictive variance—which is important for evaluating precision of predictions and is often used in stopping criteria—is also evaluated.

\end{abstract}

\begin{keyword}
Simulation \sep Gaussian processes \sep stochastic kriging \sep metamodels \sep computer experiments
\end{keyword}
\end{frontmatter}




\section{Introduction}

When computer simulation models are used to study complex systems, it is often useful to fit an empirical mathematical model to quickly approximate the time-consuming computer simulation at input values that have not yet been evaluated.  If fit to sufficient accuracy, these metamodels can replace the original computer models in optimization or ``what if" analyses. Gaussian process (GP) modeling is commonly used for fitting metamodels in simulation experiments since it provides a flexible model and model-based estimate of prediction error even if the simulation itself is deterministic.
Gaussian process models can also be used when the simulation is stochastic, although this requires an extension of the model.
Gaussian process models have become a large part in the expanding machine learning toolbox  \citep{GPMLBook}.

In this paper we are concerned with how GP models are used by practitioners, so we compare the performance of some commonly used software packages. Many practitioners are not familiar with the particulars of GP fitting, so we investigate packages that are relatively easy to use and do not require extensive knowledge of all the options and parameters that can be specified.
GP fitting is unlike linear regression where, for a given data set, all software packages will produce exactly the same parameter estimates and fitted surface (up to round-off error). Most GP fitting packages use essentially the same equations, but there is variability in how parameters are defined and estimated through numerical optimization. So, in practice, different packages can give substantially different results.  Since GP fitting is often used over other fitting techniques because of its model-based estimate of prediction error, the quality of a software implementation depends not only on the accuracy of the fitted surface, but also the accuracy of the error predictions. In many applications, random noise or computational limitations do not allow for extrinsic measures of prediction accuracy, and thus an easily obtained estimate of the uncertainty of prediction is valuable if it is at least reasonably accurate.

\subsection{Motivation}

Our major interest in GP fitting is its use to sequentially build an accurate metamodel of a computer simulation over a broad range of input values.
This global metamodel could be built automatically using excess computing capacity and then when the need for real-time decision making arises, the quickly-evaluated metamodel can be used in place of the time-consuming computer simulation model.

Sequential algorithms require a stopping criterion.
For building an accurate global metamodel, a stopping criterion that is a function of the estimated prediction error makes sense.
Many fitting methods, such as splines and neural networks, provide no estimate of prediction error apart from extrinsic methods like cross validation which are not related to the fitted model itself.  This is where the model-based estimate of GP fitting is very attractive.  In addition to its use as a stopping criterion, prediction error estimates can be used by a sequential algorithm to determine the location of the next set of design points to run with the actual computer simulation model.  We think of the sequential selection of these design points as a sequential experiment design.  A sequential experiment design is called non-adaptive if the location of each additional design point is related only to the location of the previous design points in the input space.  An adaptive sequential experiment design uses not only the location of the previous design points, but also the observed value of the response at those design points.
If the objective is to find the location of an optimum,
adaptive designs are vastly more efficient because they can select locations where the response is likely to be desirable.
However, for building an accurate global metamodel, the response information can be used to select design points where the prediction error is estimated to be large or locations that would help estimate the metamodel parameters more accurately.  Our goal is to determine which software packages have the capability of effectively and reliably estimating locations for new design points by estimating the metamodel and its prediction error.

GP metamodels are well suited for fitting an accurate global metamodel since they can fit complicated surfaces, however they are computationally slow for large data sets.
On the other hand, if the goal is to find a optimum, then a GP can also be used as a surrogate model to search for extrema \citep[e.g.,][]{jones1998efficient}.

\subsection{Emerging/growing usage of GP in the simulation community}

Many practitioners in the simulation community use Gaussian processes as a simple fitting model approach and often are not familiar with the intricacies of the model. These scientists often use the basic kriging model and do not delve into the advanced parameter settings, optimization routines, and alternative correlation models that are available.

For example, in aerospace design, \citet{christen2014global} use GPs to model the acoustic transmission on launchers in an effort to reduce damage to the payload. The GP model allows them to perform global sensitivity analysis to see which parameters in their acoustic model affect the transmission.
\citet{yin2014multiobjective} use these models in materials science for modeling functionally graded foam-filled tapered tubes to see which designs have the best energy absorption characteristics.
\citet{du2014surrogate} model the  current density  of lithium-ion batteries as a function of  eight input parameters.
GPs are used for metamodeling in simulations for corn crops by \citet{villa2012comparison}; their metamodels predict the ``nitrogen dioxide ($N_2O$) fluxes and nitrogen leaching from European farmlands."
\citet{gidaris2015performance} use kriging for earthquake engineering to see how the configuration of fluid viscous dampers  affects costs.
GPs are used as metamodels in sensitivity analysis for traffic simulation models by \citet{ciuffo2013gaussian}.

These examples highlight the need for Gaussian process software to be stable and reliable, in the same way that regression modeling is trusted for fitting linear models.
Of particular importance for these applications are reliable predictions and accurate error estimates. The error predictions are especially useful in determining whether more data is needed.

\subsection{Software discrepancies}
This study is inspired by our previous research where we found discrepancies between two software packages that were using the same GP model but were giving different results, particularly in the estimation of the prediction error. We believe that others may also encounter similar issues with GP fitting and would benefit from an in-depth study of the various software options. Knowledgeable users may know how to improve the results by setting advanced options or tuning parameters. However, we are trying to find what works best for practitioners who may not have this specific knowledge. Thus, we select packages that are easy to use, and we have left as many options to the default setting as possible.  For our comparisons we use packages from a mixture of platforms: the R \citep{r-core-team} packages  DiceKriging, GPfit, laGP, and mlegp;
JMP, produced by SAS;
the MATLAB toolbox DACE;
and the Python modules GPy and sklearn (scikit-learn);
These are described in more detail in Section \ref{sectionpackagedetails}.
JMP is a commercial package, the rest are free and publicly available.

\begin{figure}[ht]
\includegraphics[width=\textwidth]{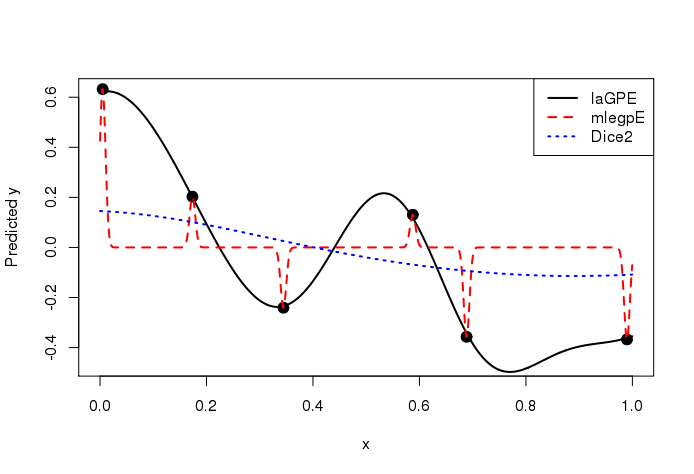}
\caption{
Comparison of Gaussian process fits from three software packages, laGPE, mlegpE, and Dice2, on one-dimensional data.
The black points are the input/output data given to each package to fit a GP model.
The lines are the predicted mean over the interval [0,1] for each package, showing significant differences.
}
\label{RGPP_D1_B1.3_SS6}
\end{figure}

Figure \ref{RGPP_D1_B1.3_SS6} is a simple example that demonstrates problems that can arise.
It displays a sample of size six (black points) in one dimension fit using common options for three different software packages, which then give predictions of the output for $x$ between 0 and 1.
The details of the packages will be explained later in this paper.
The predictions given by laGPE smoothly interpolate between the observations.
We can see that mlegpE exhibits mean reversion, where it predicts the observed points correctly, but then quickly reverts to the mean away from those points.
At the other extreme, Dice2 oversmooths, causing its predictions to be far from the observed points.
Furthermore, the predictions of the standard error across the range of $x$ values will also be significantly different.
Thus, even for a simple data set, we can obtain very different results.

\section{GP fitting}

\subsection{Model}

A Gaussian process is characterized such that the output from any set of input points has a multivariate normal distribution. If we have $n$ inputs in $d$ dimensions, then the $i^{th}$ input is
$ \bm{x_i}=(x_{i1} , ... , x_{id})^T$. These are stored in the rows of the  $n$ by $d$ input matrix $\bm{X}$.
The output is one dimensional $\bm{y} = (y_1, ..., y_n)^T$.

Following \citet{Sacks1989} the surface is modeled as a mean, $\mu$, plus a Gaussian process, $z$, which is a function of $\bm{x}$ as follows:

$$ \displaystyle y = f(\bm{x}) = \mu + z(\bm{x}) . $$

In general, a linear combination of functions can be used in place of $\mu$, which is called universal kriging (e.g., \citet{bastos2009diagnostics}).
Many authors use  only a constant term for $\mu$, since the Gaussian process is flexible enough to model any linear behavior in addition to many other more interesting features of the surface.
In this paper we also use a mean-only model, following the suggestion of \citet{chen2016analysis}, who claim that ``there is little to be gained (and maybe even something to lose) by using other than a constant term for $\mu$.''
One explanation is that replacing this constant by a more complicated function $f(\bm{x})$ (e.g., a polynomial of an order higher than zero) requires the estimation of additional (often extraneous) parameters (e.g., polynomial coefficients).

Under this model, the distribution of the outputs, $\bm{y}$, is multivariate normal with mean $\mu \bm{1_n}$, where $\bm{1}_n$ is the $n$-length vector of ones.  The covariance matrix of this multivariate normal distribution is proportional to a correlation matrix, which has a special structure such that the points will create a smooth surface.  The constant of proportionality between the covariance matrix and the correlation matrix is a variance that is denoted $\sigma^2$. The correlation matrix is constructed such that the correlation of the outputs from any two distinct points in the input domain is inversely related to the distance between those two points.
In particular, the correlation goes to one as the distance between the two input points goes to zero, and the correlation goes to zero as the distance goes to infinity.  This allows the output points to form a surface, but also places few constraints on the shapes and features of that surface.
Thus the model is

\begin{equation} \label{model1eqn}
    \bm{y} \sim  MVN\left( \mu\bm{1}_n, \sigma^2 \bm{R} \right) ,
\end{equation}

\noindent where $\bm{R}$ is the correlation matrix of $\bm{y}$. $R_{ij}$ denotes the element in the $i$th row and the $j$th column of $\bm{R}$, and is the correlation between $y_i$ and $y_j$.

The covariance matrix is determined by the correlation function, for which there are many options. The most common covariance function, which we use here, is the Gaussian correlation that defines the correlation between the outputs at $\bm{x_i}$ and $\bm{x_j}$ as

\begin{equation} \label{gaussiancorrelation}
    R_{ij} = \prod^d_{k=1} \exp{\left( -\theta_k \left( x_{ik} - x_{jk} \right)^2  \right)}.
\end{equation}

\noindent The estimator for the mean is

$$ \hat{\mu}  = \left( \bm{1}_n \bm{R}^{-1} \bm{1}_n \right)^{-1} \left( \bm{1}_n \bm{R}^{-1}\bm{y} \right).$$

\noindent Then the best linear unbiased predictor (BLUP) of $y$ at $\bm{x}$ is given by \citet{Sacks1989}.  Using our notation and following the derivations in \citet{GPfit}, we find

\begin{equation} \label{yPred}
\hat{y} = \hat{f}(\bm{x}) = \hat{\mu}  + \bm{r}^T \bm{R}^{-1} \left(\bm{y} - \hat{\mu}\bm{1}_n \right) = \left[ \frac{(1-\bm{r}^T \bm{R}^{-1}\bm{1}_n)}{\bm{1}_n^T \bm{R}^{-1} \bm{1}_n}\bm{1}_n^T + \bm{r}^T \right] \bm{R}^{-1} \bm{y} = \bm{C}^T \bm{y} ,
\end{equation}

\noindent where $\bm{r}^T = (r_1(\bm{x}), \ldots, r_n(\bm{x}))$,   $r_i(\bm{x})$ is the covariance between $\bm{x_i}$ and $\bm{x}$, and $\bm{C}$ is a vector of length $n$ defined as shown. The associated mean squared error of $\hat{y}$ at $\bm{x}$ is

\begin{equation} \label{MSEPred}
\begin{split}
 \varphi (\bm{x}) = \sigma^2 \left[1 - 2\bm{C}^T \bm{r} + C^T \bm{R} \bm{C} \right] = \sigma^2 \left[1 - \bm{r}^T \bm{R}^{-1} \bm{r} +  \frac{(1-\bm{1}_n^T \bm{R}^{-1} \bm{r})^2}{\bm{1}_n \bm{R}^{-1} \bm{1}_n} \right]  .
\end{split}
\end{equation}

These parameters can be estimated, and then used in equations \ref{yPred} and \ref{MSEPred} to get the predictions.
The error added from parameter estimation is generally not included in the predictive equations, but can be estimated through bootstrap techniques, see \citet{kleijnen2015design}.

\subsection{Nugget effect}

The model of equation \ref{model1eqn} does not account for random noise. This can be done by adding a nugget parameter.
To account for random homoscedastic (constant variance) noise, the model above must be augmented as follows:

\begin{equation*}
\bm{y} \sim  MVN\left( \mu\bm{1}_n, \sigma^2 (\bm{R} + \delta \bm{I})\right) .
\end{equation*}

If a nugget, $\delta$, is included in the model, then the correlation matrix is increased along the diagonal by $\delta$:

$$ \bm{R_{\delta}} = \bm{R} + \delta \bm{I} . $$

\noindent
Then the predictive equations \ref{yPred} and \ref{MSEPred} can be used with $\bm{R_{\delta}}$ in place of $\bm{R}$. Note that $\bm{R_{\delta}}$ is no longer a correlation matrix since the diagonal values are greater than one.

The nugget has the effect of smoothing the function and allowing for noise.
Another reason for using a nugget is to provide computational stability. The calculations above all require inverting $\bm{R}$, which can be near singular. Adding a nugget will improve this stability.
When the noise is heteroscedastic (i.e. the noise variance varies across the input domain) then stochastic kriging methods must be used as explained in Section \ref{sectionstoch}.

\subsection{Parameters}\label{subsectionparams}

When fitting a model to data, at least $d + 2$ parameters must be estimated: the mean $\mu$, the $d$ correlation parameters $\bm{\theta}$, and the $\sigma^2$ parameter. One additional parameter, $\delta$, must also be estimated if the nugget is used.
Given the same parameters and the same data, different software should give the same predictions since they are using the same equations. Thus, the differences that we have observed in predicted values between software packages are likely caused by different parameter estimates.
Each software package uses some type of numerical optimization method to seek estimators for these parameters that maximize the likelihood function, see Section \ref{sectionoptimizationtechniques}.
Practitioners typically trust these optimization methods to work without intervention. Later in this paper, empirical studies essentially demonstrate the performance of the optimization methods for various software packages.

The mean for the Gaussian process is $\mu$. Predictions far away from design points will revert to the mean since they will have low correlation with the observed data, but $\mu$ has a much smaller effect on predictions near design points.
As previously discussed, the mean term can be replaced with a linear model-type function $\bm{\beta}^T\bm{f}$, but we do not consider such an expanded model in this paper.
Some analysts prefer to not use a mean term at all.

The vector of hyperparameters, $\bm{\theta} = (\theta_1, ..., \theta_d)^T$, contains the correlation parameters of the covariance function. There is one parameter for each dimension that determines how strong the correlation is between points in corresponding dimension.
Sometimes a different parameterization is used
(see  Section \ref{corrsubsection}),
which can lead to changes in the ease and stability of numerical optimization.

The parameters in $\bm{\theta}$ help determine the correlation, but the parameter $\sigma^2$  also affects the fitting since the covariance between any two points is $cov(y_i,y_j) = \sigma^2 R_{ij}$. Note that this variance parameter is not the variance of a sample from the output surface. By itself it can be interpreted as the variance of a point ``infinitely far" from all other points.

The nugget allows for measurement error or stochasticity of the response. If the nugget is not used (i.e., set to zero), then the model will interpolate exactly, so the prediction error at a design point will be zero. This is often useful for deterministic computer experiments, but if the data is stochastic then a nugget should be estimated and used.
Using a nugget improves the numerical stability by making the correlation matrices easier to invert; this inversion can be a problem when there is a large number of sample points.
\citet{ranjan2011computationally} provide a method to use the smallest nugget value that makes the computations stable, which can help balance the benefits of having stability and a small nugget. \citet{gramacy2012cases} argue that the nugget provides protection when the assumption of stationarity is violated or the data is sparse, and they claim this protection is more important than stability.

\section{Gaussian process fitting in various software packages}
\label{sectionpackagedetails}

Table \ref{packagetable} lists the primary sources for our comparison of packages.  Most of the packages provide users with options.
We try to leave most options at their default settings, since that is what most practitioners are likely to use. However, in order to have fair comparisons, we make some simple selections so all packages use comparable models.
This section provides overviews of the packages and describes the selections we use in our study.

\begin{table}[ht]
\centering
\caption{Packages we are using}
\label{packagetable}
\begin{tabular}{|l|l|l|l|}
\hline
\textbf{Package}  & \textbf{Version}    & \textbf{Platform} & \textbf{Primary source} \\ \hline
DiceKriging     & 1.5.5   & R         &      \citet{roustant2012dicekriging}          \\ \hline
GPfit     &  1.0-0   & R         &      \citet{GPfit}          \\ \hline
laGP       & 1.3-2   & R        & \citet{laGP:cran} \\  \hline
mlegp      & 3.1.4  & R         &      \citet{dancik2008mlegp}          \\  \hline
JMP     & Pro 13.0.0   & JMP         &      \citet{JMP-GP}          \\ \hline
DACE     & 2.5    & Matlab     &  \citet{lophaven2002dace}              \\ \hline
GPy      & 1.5.6    & Python     &   \citet{GPy}        \\ \hline
sklearn &  0.18.1  & Python      &  \citet{pedregosa2011scikit}           \\ \hline

\end{tabular}
\end{table}

\textbf{DiceKriging} is an R package for kriging created by the DICE Consortium, which has also released the R packages DiceOptim, DiceDesign, and DiceEval \citep{roustant2012dicekriging}. DiceOptim does the optimization performed when fitting DiceKriging models.
DiceKriging provides many options for fitting and is very thorough, so it may be a good choice for many R users.

\textbf{GPfit}, another R package, was created by MacDonald, Ranjan, and Chipman.
GPfit does the most extensive search in optimizing the maximum likelihood parameters, as detailed in \citet{GPfit}. Even when the control parameters for the optimization were set to reduced values, GPfit was still  slower than the other packages by orders of magnitude. With more than a hundred design points GPfit becomes prohibitively slow, while the other packages still run quickly. One advantage of GPfit is that it focuses on computational stability, using the ideas put forth in \citet{ranjan2011computationally}. It sets the nugget to be the smallest value that will avoid singularity, meaning that the nugget is never estimated.  Thus GPfit is best suited for noiseless data.
Since the default exponential power is 1.95 (instead of 2; see Equation \ref{gaussiancorrelation}), we include it as a separate model in some of the testing below.
However, we change the power to 2 for most of the testing to be comparable with the other packages.

\textbf{laGP} is an R package created by Robert Gramacy that provides an entirely new method for fitting using GPs \citep{laGP:cran}. The ``la" stands for ``local approximate" since the model is designed for large data sets. In the laGP model, sparsity is exploited so the kriging is done on a small number of design points that are most important for prediction at a given point \citep{BG:LGPA}. Thus, it can be run when there are a million design points, since it uses only a small number of points to make predictions and leverages parallelism. For the purpose of this research, we only use the basic GP fitting functions provided by the package. Although it is an R package, the heart of it is a C implementation wrapped in R, meaning it should be faster than a basic R implementation. In its current state it is a rather minimal package with a focus on speed, so it has fewer additional options and may require more fine-tuning---but is extremely fast.

The R package \textbf{mlegp} (Maximum Likelihood Estimates of Gaussian Processes) was created by Garrett Dancik \citep[see][]{mlegp:cran}. It provides full GP modeling capabilities. A distinctive feature of mlegp is that the user can specify the nugget matrix up to a multiplicative constant, which can be useful when the response is heteroscedastic as in \citet{dancik2008mlegp}.
Another feature is the ability to perform sensitivity analysis, letting the user quantify how the response is affected by parameters and how much variability in the output can be attributed to changes in the design matrix.

We also use the Gaussian Process capability of \textbf{JMP}, a data analysis software tool provided by SAS \citep{JMP-SAS}. This program is commonly used by practitioners since it provides a clean interface, makes data analysis simple, and provides useful output displays.

\textbf{DACE} is a Matlab toolbox for fitting data from deterministic computer experiments, so it does not allow for noisy data  \citep[see][]{lophaven2002dace}.
Thus it is only suitable when there is no random error at any given design point.
DACE was created and last updated in 2002, so while it is commonly used, it lacks many of the additional features that other packages include.

\textbf{GPy} is a Python Gaussian process implementation created by the Sheffield machine learning group \citep{GPy}. GPy has a tremendous amount of functionality available for many different cases. During our preliminary tests, GPy gave poor results due to computation problems when it was version 0.6.  These problems were fixed after GPy version 1.0 was released in April 2016, and we report results for version 1.5.6 in this paper.

Another open source library for Python is scikit-learn, which we call \textbf{sklearn} since that is the name of the Python module \citep{pedregosa2011scikit}. It is targeted for machine learning, not just kriging, so there are many other modeling options available in the module.
Up through version 0.17, the kriging implementation was based on DACE. However, the Gaussian process functionality was vastly upgraded with version 0.18, released in September 2016 \citep{scikitlearnReleaseHistory}.
The update added options for the correlation function, called the kernel, including the Gaussian, 
Mat\'{e}rn, rational quadratic kernel, and others, as well as sum or product combinations of kernels.

Since they all use nearly the same equations, the real challenge in model fitting is estimating the parameters. Whereas the predictions are calculated using formulas, parameters must be estimated by solving an optimization problem. Generally, the parameters are chosen to be those that maximize the likelihood. However, the solution found to this depends on starting values, bounds, the algorithm used, and the parameterization.
There is additional ambiguity since real data is typically not truly samples from an actual Gaussian process. The Gaussian process model is just a useful approximation technique, and thus there are no true parameter values.
The rest of this section examines the differences between
the package parameterizations,
the options available,
and the estimation methods.


\subsection{Correlation functions and parameterizations}
\label{corrsubsection}

The commonly used Gaussian correlation function is shown in Equation \ref{gaussiancorrelation} above, but there are many variations on it and many other correlation functions that can be used. \textbf{DACE} and \textbf{JMP}  use this standard formulation.
The \textbf{mlegp} package uses a different notation for the correlation parameters, $\bm{\beta} = \bm{\theta}$, but this does not affect the calculations at all.
\textbf{GPfit} uses $\bm{\beta} = \log_{10} \bm{\theta}$ in order to
focus the optimization search near the center of the search space.

The Gaussian correlation can be generalized by allowing the exponent to be changed to any value in the range $[1,2]$, which allows for different smoothness in the surface.
The default correlation function for \textbf{GPfit} uses 1.95 in the exponent.
\citet{ranjan2011computationally} justifies this change by explaining that it helps to reduce the computational problems caused by a near-singular correlation matrix when a space-filling design is used.
In our tests, we evaluate two versions of GPfit: one using the Gaussian correlation function, which we call GPfit2, and one using 1.95 as the exponent, which we call GPfit1.95.

The package \textbf{laGP} moves $\theta$ to the denominator, and calls it the length-scale parameter, as shown in Equation \ref{gaussiancorrelationlaGPCOPIED}.
This change is simply a reparameterization and does not affect the model at all. However, it will affect the optimization routine used to estimate $\bm{d}$, the $p$-length vector where the $k$th element is $d_k$.
This formulation is used by other authors such as \citet{GPMLBook}. The correlation function is
\begin{equation} \label{gaussiancorrelationlaGPCOPIED}
    R_{ij} = \prod^p_{k=1} \exp{\left( -\left( x_{ik} - x_{jk} \right)^2 / d_k   \right)},
\end{equation}
where $p$ is used to denote the number of dimensions. The notation used for the parameters is also slightly different in the laGP code and vignette (a guide for the R package) than the others. The nugget is referred to as $g$ in the code and $\eta$ in the vignette, while the lengthscale parameters are denoted by $\bm{d}$ in the code and $\bm{\theta}$ in the vignette \citep{laGPvignette}.

Another parameterization adjusts the correlation function so that the lengthscale parameters, denoted as $\bm{\ell}$, appear squared in the denominator.
This puts $\bm{\ell}$ on the same scale as $\bm{x}$.
In addition a factor of two is added in the denominator so that the correlation function closely resembles the Gaussian probability distribution function, as shown in Equation \ref{gaussiancorrelationGPyCOPIED}.
This is used by \textbf{DiceKriging}, \textbf{sklearn} and \textbf{GPy} \citep{GPyDocumentation}.
\begin{equation} \label{gaussiancorrelationGPyCOPIED}
    R_{ij} = \prod^p_{k=1} \exp{\left( - \frac{1}{2} \left( x_{ik} - x_{jk} \right)^2 / \ell_k^2   \right)}.
\end{equation}

Another popular correlation function is the Mat\'{e}rn function.
It takes a parameter $\nu$ that determines the smoothness. Commonly used values for $\nu$ are 3/2 and 5/2.
The Mat\'{e}rn correlation function can be seen as a generalization of the Gaussian correlation function since they are equivalent for $\nu= \infty$.
 According to \citet{roustant2012dicekriging}, the default correlation function for \textbf{DiceKriging} is the Mat\'{e}rn with $\nu=5/2$, which is

\begin{equation}  \label{maternequation}
\setlength{\jot}{10pt}
\begin{split}
    g(h) &= \left( 1 + \sqrt{5}|h| + \frac{5}{3}h^2 \right) \exp \left( - \sqrt{5}|h|\right), \\
    \text{where } h &= \sqrt{\sum_{k=1}^{p} \left( x_{ik} - x_{jk} \right)^2 / \ell_k^2}.
\end{split}
\end{equation}

GPfit, GPy, and sklearn have the Mat\'{e}rn correlation as an option.
We include two versions of DiceKriging in our comparisons, one using the Gaussian correlation, which we label Dice2, and another using the Matérn $\nu=5/2$ correlation function, since it is the default for DiceKriging, which we label DiceM52.

\subsection{Nugget options}
\label{nuggetsubsection}

There are also options available for the nugget parameter.

\textbf{DiceKriging} defaults to having no nugget. There is also the option of setting the nugget to a constant or estimating it.
For both DiceKriging with the Gaussian (Dice2) and the Mat\'{e}rn (DiceM52), we let it estimate the nugget. Preliminary tests involving noiseless data did not reveal noticeable differences between fits based on no nuggest and an estimated nugget. We chose to include the version that estimates the nugget since DiceKriging is also useful for stochastic kriging.  See more details in Section \ref{sectionstoch}.

\textbf{GPfit} uses the smallest nugget value that keeps the computation stable, as explained in \citet{ranjan2011computationally}, in order to prevent over-smoothing. The nugget value they use is
$$\delta_{lb} = \max \left\{ \frac{\lambda_n \left( \kappa \left( \bm{R} \right) - e^a \right)}{\kappa\left(\bm{R}\right)\left(e^a - 1\right)} , 0 \right\},$$
\noindent where $\lambda_n$ is the largest eigenvalue of $\bm{R}$, $a$ is a parameter set to 25 for space-filling designs, and $\kappa(\bm{R})$ is the condition number of $\bm{R}$.  \citet{GPfit} compare GPfit to mlegp and state that
``mlegp occasionally crashes due to near-singularity of the spatial correlation matrix,"
which agrees with what we have seen, so there is a benefit to setting the nugget in this way.

In \textbf{laGP}  the user must either set the nugget to a fixed value or tell laGP to estimate it, since there is no default option. We tried several values for the nugget in preliminary tests, and found that $1 \times 10^{-6}$ worked best. Thus, in our tests we run laGP both with the nugget set to $1\times 10^{-6}$, called laGP6, and with the nugget being estimated, called laGPE.
laGP is also different from the other packages we use since it performs the calculations from a Bayesian perspective.
In practice this makes little difference for users, since the default priors are very general.
In addition, laGP is the only package we use that does not estimate a mean term.

By default, \textbf{mlegp} will not use a nugget unless there are repeated design points, but it can be estimated or set to a constant or a vector \citep{dancik2011mlegp}. For our tests we run both with nugget fixed to 0, and with a nugget estimated using a starting value of $1 \times 10^{-6}$. We call these mlegp0 and mlegpE, respectively. As shown below, we find there is little difference.

\textbf{JMP} provides an option to fit the model with no nugget or with an estimated nugget.
We run JMP using the Gaussian correlation both with estimating a nugget and without a nugget, and we refer to these two as JMPE and JMP0, respectively.

Since it is designed for noiseless computer experiments, \textbf{DACE} does not let the user set or estimate the nugget. Instead it uses a small value equal to $2.22 (10+n) \times 10^{-16}$  for computationally stability.

The nugget can be set or estimated in \textbf{GPy}
by setting the noise variance parameter when using the GPRegression function.
We set this parameter to a small value, $1 \times 10^{-8}$, which forces the model to estimate a nugget parameter.

The nugget is called alpha by \textbf{sklearn}, and defaults to $1 \times 10^{-10}$.
Alternatively, the nugget can be specified by using a WhiteKernel, and this method allows it to be estimated. We use the default value in our investigation.

Table \ref{nuggettable} shows the packages we use in our study, along with  the differences in the parameterization of $\bm{\theta}$ and the options and defaults for the nugget. The last column shows how we set the nugget for our study.

\begin{table}[ht]
\centering
\caption{Settings for each software version used in the study. DiceM52 uses the Mat\'{e}rn correlation function with $\nu=5/2$, GPfit1.95 uses the power exponential correlation with power 1.95, and all other use the Gaussian correlation function.}
\label{nuggettable}
\begin{tabular}{|l|c|c|c|c|c|}
\hline
\multirow{3}{*}{\textbf{Package}} & \multirow{3}{*}{$\bm{\theta}$} & \multicolumn{4}{c|}{\textbf{Nugget}}      \\ \cline{3-6}
&  & \textbf{Can}   & \textbf{Can} & \textbf{}  & \textbf{Setting } \\
&  & \textbf{set?}   & \textbf{estimate?} & \textbf{Default}  & \textbf{ used} \\ \hline

\begin{tabular}{l} DiceM52 \\  Dice2 \end{tabular}                               & $\bm{\theta^2} = 1/(2\bm{\theta})$      & \checkmark          & \checkmark             & 0        & Est                                    \\ \hline

\begin{tabular}{l} GPfit1.95 \\  GPfit2 \end{tabular}                               & $\bm{\beta} = \log_{10}\bm{\theta}$      & -          & -             & $\delta_{lb}$        & $\delta_{lb}$                                    \\ \hline

\begin{tabular}{l} laGP6 \\  laGPE\end{tabular}                                & $\bm{d} = 1/\bm{\theta}$             & \checkmark & \checkmark    & NA                   & \begin{tabular}{c} 1e-6 \\  Est\end{tabular} \\ \hline

\begin{tabular}{l} mlegp0 \\ mlegpE\end{tabular}                               & $\bm{\beta} = \bm{\theta}$               & \checkmark & \checkmark        & 0  & \begin{tabular}{c} 0 \\  Est\end{tabular}                                                \\ \hline
\begin{tabular}{l} JMP0 \\  JMPE\end{tabular}                                & $\bm{\theta} = \bm{\theta}$              & \checkmark & \checkmark    & NA                   & \begin{tabular}{c} 0 \\  Est\end{tabular} \\ \hline

\begin{tabular}{l}  \end{tabular} & &  & & &   \\[-10pt]
\begin{tabular}{l} DACE \end{tabular}                                & $\bm{\theta} = \bm{\theta}$              & -          & -             & (10+n)2.2e-16                    & (10+n)2.2e-16                                                \\[2pt] \hline

\begin{tabular}{l}  \end{tabular} & &  & & &   \\[-10pt]
\begin{tabular}{l} GPy  \end{tabular}           &  $\bm{\ell}^2 = 1/(2\bm{\theta})$             & \checkmark &      \checkmark   & Est                    &           Est                                       \\[2pt] \hline

\begin{tabular}{l}  \end{tabular} & &  & & &   \\[-10pt]
\begin{tabular}{l} sklearn \end{tabular}                        & $\bm{\ell^2} = 1/(2\bm{\theta})$              & \checkmark & \checkmark              & 1e-10              & 1e-10                                          \\[2pt] \hline

\end{tabular}
\end{table}

\subsection{Optimization techniques}\label{sectionoptimizationtechniques}

There also many options in most packages for setting up the optimization routine that estimates the parameters. There are different algorithms, choices of starting points and number of restarts, and stopping criteria.
We use the default settings for all packages unless specified otherwise.

By default, \textbf{DiceKriging} uses the L-BFGS-B algorithm for parameter optimization. The other option available, but not used in our study, is the genoud algorithm from the rgenoud R package \citep{mebane2011genetic}, that combines genetic (evolutionary search) algorithms with derivative-based algorithms.

\textbf{GPfit} uses the most in-depth optimization algorithm. As detailed in \citet{GPfit}, GPfit uses L-BFGS-B \citep{byrd1995limited} with multiple starts to estimate the correlation parameters which they have transformed to be $\bm{\beta}=\log_{10} \bm{\theta}$.  This transformation focuses the optimization search more to the middle of the search space than to the edges.
In Section 2.3 of \citet{GPfit}, they describe how GPfit adds bounds for each $\beta_k$ to create a domain where the optimum is likely to be found.
The function that is minimized is the negative profile log-likelihood (which they call the deviance),
$$-2\log L_{\bm{\theta}} \propto \log |\bm{R}| + n \log\left[\left(\bm{y} - \bm{1}_n \hat{\mu}(\bm{\theta}) \right)^T \bm{R}^{-1} \left((\bm{y} - \bm{1}_n \hat{\mu}(\bm{\theta}) \right)\right] . $$
GPfit begins its search with a space-filling LHD in the space of all the $\beta_{i}$'s, then selects a number of parameter sets that have low deviance. These points are clustered using the $k$-means algorithm. Then the L-BFGS-B algorithm is run using these cluster centers as the starting point in each restart.

\textbf{laGP} requires an initial value for the correlation parameters and nugget (if estimated) with no default provided. However laGP provides the functions \code{darg} and \code{garg} which provide good initial starting values for $\bm{\theta}$ and $\delta$ using Empirical Bayes \citep{laGP:cran}. We use these two functions in our tests to find starting values.

The package \textbf{mlegp} estimates the parameters using L-BFGS \citep{liu1989limited} in a gradient method \citep{mlegp:cran}. The starting points are found using multiple Nelder-Mead simplexes.

\textbf{JMP} is proprietary software and provides no details on the optimization or other details beneath the surface.
There are no options that can be set for the optimization.

\textbf{DACE} uses a pattern search that iterates through the steps of exploring, moving, and rotating, after finding a suitable starting point as in \citet{lophaven2002aspects}.
By default, DACE uses a single correlation parameter for all dimensions and initializes it to 0.1. In order to be comparable to the other methods, we retain the 0.1 value, but make it into a $d$-length vector so that these packages fit a correlation parameter for each dimension. In DACE, upper and lower bounds for each $\theta_i$ must be provided. We set these to be generous, with the lower bounds to $1\times 10^{-4}$ and the upper bounds to $1 \times 10^{3}$.

\textbf{GPy} begins with initial correlation parameters set to 1. However, by default GPy uses the same correlation parameter in every direction, so to get a separate parameter for each dimension, we had to set \code{ARD=True}. We also had to set the likelihood variance to a small value of $1\times 10^{-8}$, instead of its default of 1, to get good results.
GPy allows a choice of optimization routines: TNS, L-BFGS-B, and BFGS from Scipy \citep{scipy}; Adadelta and RProp from the Python module climin; as well as Nelder-Mead simplex routine and Scaled Conjugate Gradients.
The optimization is run through the Python module
``\code{paramz}."
We use the default of L-BFGS-B.
We also use five optimization restarts to ensure the optimization results are favorable, although this increases the run time.

\begin{sloppypar} 
By default, \textbf{sklearn} uses the ``\code{fmin\textunderscore l\textunderscore bfgs\textunderscore b}" optimization algorithm from scipy.optimize \citep{scikitGPRegressordocumentation,scipy}. This algorithm is an implementation of the L-BFGS-B algorithm \citep{byrd1995limited}.
There is an option to use multiple restarts to help the optimization avoid getting stuck in a local minima.
By default the number of restarts is zero,
which is what we use in our tests.
However, trying more restarts may improve performance. 
In our initial testing with sklearn, we observed poor results when the data was not scaled.
In our comparison tests in this paper, we scaled all our data to have mean 0 and range 1, as discussed in Section \mbox{\ref{empiricalsection}}.
\end{sloppypar}

\section{Empirical study methodology}\label{esmethod}

In this section, we discuss the criterion on which we will compare different GP fitting software packages.
When constructing a global metamodel, two properties of GP modeling are of primary importance:
(1) the accuracy of prediction, and (2) the accuracy of the estimate of prediction error.  The first is important for obvious reasons.  The second is important to allow the practitioner to assess whether the metamodel is fit for use or whether additional data is needed to improve its fit
The differences between parameterizations mentioned in Section \ref{corrsubsection} do not matter here because our interest is on the accuracy of the predictions.
We focus on global fitting, not on optimization where the comparison criterion would be the accuracy of the estimation of the optimal input vector and the estimated output scalar.

\subsection{Model accuracy}

When we evaluate model accuracy, we use a known surface and compare the actual surface and the model's predicted values at a large number of points, called prediction points. The prediction points are distributed throughout the area of interest for the input values.  We use the square root of the mean of the squared errors at the prediction points as an estimate of the model's RMSE; for ease of discussion we will call this estimate the ``empirical model RMSE" or ``EMRMSE." Thus, using $m$ prediction points $\bm{x}_1^*$, $\cdots$, $\bm{x}_m^*$,

$$ \text{EMRMSE} = \sqrt{\frac{1}{m} \sum_{i=1}^m (\hat{y}(\bm{x}_i^*) - y(\bm{x}_i^*))^2}. $$

\noindent  \citet[][p. 108]{santner2003design} call this the empirical root mean squared prediction error. 
Although this metric does not account for the distribution of the prediction errors, it is a commonly-used single number that summarizes the quality of the fit of the model across the entire surface.

We use EMRMSE to assess the quality of the fit for a model. This measure is primarily useful when comparing two models fitting the same surface since the model with the lower EMRMSE fits the surface better.
Also, in our empirical studies in Section \ref{empiricalsection}, the value of EMRMSE can be very roughly thought of as an average relative error for the model.
This is because EMRMSE estimates the standard deviation of the prediction errors, and, as discussed in Section \ref{empiricalsection}, we scale the responses (at prediction and design points) to have a range of 1. 

\subsection{Accuracy of estimated prediction error}

To evaluate the accuracy of the model's estimated prediction error,
we estimate the model's mean squared error $\varphi(\bm{x})$, defined in Equation \ref{MSEPred}, by
$\hat{\varphi}(\bm{x})$, obtained by substituting the fitted model's parameter estimates for the unknown parameters in
equation (\ref{MSEPred}).
The square root of the average of the estimated mean squared errors over all prediction points is used as the summary measure for the predicted model RMSE and called the ``PMRMSE."

$$ \text{PMRMSE} = \sqrt{\frac{1}{m} \sum_{i=1}^m \hat{\varphi}(\bm{x}_i^*)}$$


Since EMRMSE and PMRMSE both measure the model's RMSE, we expect them to be approximately equal.
If we observe EMRMSE $\approx$ PMRMSE, that confirms the accuracy of the model's prediction error.
If EMRMSE is much larger than PMRMSE, then the model is overconfident in its fit, since its estimated prediction errors will be smaller than the empirical errors.
Conversely, if EMRMSE is much less than PMRMSE, then the model's estimated prediction errors are conservative.

Note that \citet{bastos2009diagnostics} suggest a different way to compare predictive errors by using the predictive covariance matrix.  For a single prediction point, we could calculate the standardized prediction error, which should follow a $t$-distribution. For the entire set of prediction points, these will be correlated. Intuitively this makes sense because the prediction functions are continuous and the true surface is usually also continuous, so points near each other will necessarily have related errors.  This error analysis requires the predictive covariance matrix so that the standardized errors can be decorrelated.  This method places equal importance on all parts of the surface. However, in practice, one usually focuses on areas where the predicted error is large.  Moreover, most software packages do not provide the predictive covariance matrix, and the errors can differ by orders of magnitude depending on how close they are to sample points.
For these reasons, we use EMRMSE and PMRMSE as the basis for assessing the accuracy of the estimated prediction error for a given model.

\subsection{Comparison to Linear Model}

Fitting GPs can be computationally intensive when the number of points is large. Thus we only want to make the computational investment when we will see a significant improvement over simpler models. In preliminary investigations, we found that when the surface is too trivial or the sample size is too small that fitting a linear model---or even just the mean---can give predictions similar to the fitted GP model.
Thus, in our empirical study, whenever we fit a GP model in $d$ dimensions, we also fit a $d$-dimensional hyperplane, which we call the linear model (LM).
Using the same prediction points, we then calculate the EMRMSE for the LM.
Use of the complicated GP model is  only beneficial if the EMRMSE of the GP model is substantially less than that of the LM.
When the EMRMSE of the GP model is greater than or equal to that of the LM, it indicates that the GP model is unsuitable for the situation.
In either case, comparisons of GP model fitting are not of interest.
Thus throughout the empirical study we will define $\xi(\text{M})$ as the ratio of the EMRMSE of the fitted GP model M, and the EMRMSE of the LM, as shown in
Equation~\ref{xiequation}:

\begin{equation} \label{xiequation}
 \xi(\text{M}) = \frac{\text{EMRMSE(M)}}{\text{EMRMSE(LM)}}.
\end{equation}

\noindent This is similar to the normalized RMSE measure, $e_{\text{rmse,ho}}$, used by \citet{chen2016analysis}, which is the ratio of the RMSE of the GP model to the RMSE of the model that only fits the mean.
We believe $\xi$ is more useful because practitioners are more likely to consider a linear model as an alternative to the Gaussian process.

To keep PMRMSE on the same scale, we also define $\pi(\text{M})$ as the ratio of the PMRMSE of M to  the EMRMSE of the LM, as shown in
equation~\ref{piequation}:

\begin{equation} \label{piequation}
 \pi(\text{M}) = \frac{\text{PMRMSE(M)}}{\text{EMRMSE(LM)}}.
\end{equation}

\section{Empirical study results}
\label{empiricalsection}

In this section we compare the aforementioned software packages on four test functions:
the borehole function,
the output transformer-less (OTL) circuit function,
the Dette and Pepelyshev 8-dimensional model,
and the Morris function.
For all the functions,
we created independent maximin Latin hypercube samples (LHSs) using the R package \code{MaxPro} \citep{R-MaxPro} for the design matrices, and a much larger (2,000 point) maximin LHS for the prediction points. Using a maximin LHS helps ensure that the data represents the input space well.
We use $\bm{x_i} \in [0,1]^d$,
which is commonly done to avoid numerical issues and make sure the data scale is reasonable for the correlation function.
When calculating function values, the input values, $\bm{x_i}$, are scaled to be in the appropriate domain of each function.
The output, $\bm{y}$, can also be standardized before fitting the model to it since the range can affect how much of the variation in the data is seen as noise.
Many software packages do this standardization automatically or have the option to do so.
For all of our comparisons shown in Section \ref{empiricalsection}, the output data is scaled to have mean 0 and range 1, as recommended by \citet{gramacy2007tgp}.

A common recommendation in computer experiments is to use a sample size of ten times the number of input dimensions, i.e., choose $n=10d$ \citep{loeppky2012choosing}.
We find that this sample size is often too small, giving predictions only slightly better than a linear model.
Thus we  use input samples of size $n=10d$ and $n=20d$ taken from space-filling designs in our comparisons that follow.
The amount of data needed to get a good fit depends on the curvature of the data, the quality of the design, and the desired accuracy of the model.

On each sub-plot of the following figures, we generate five surfaces, called macroreplicates, and fit them using thirteen software package versions.
Different shaped icons represent the results for different macroreplicates.
The five macroreplicates are generated by five different sets of design and prediction points.
Thus,  we have 65 fitted metamodels on each plot---thirteen packages fitting five macroreplicates each.
The x-axis plots $\xi(\text{M}_i)$ and $\pi(\text{M}_i)$ for each macroreplicate, as defined in equations \ref{xiequation} and \ref{piequation}, where $\text{M}_i$ represents one of the 65 metamodels.
$\xi(\text{M}_i)$ is plotted on the solid line for each package with solid icons,
while the gray icons slightly above each solid line are $\pi(\text{M}_i)$.
Lines connecting the $\pi$ value to the $\xi$ value for each macroreplicate make it easy to see whether the packages are underestimating or overestimating the actual error.
A positive slope for the line indicates overestimation of the error, while a negative slope indicates underestimation of the error.
Thus, a good metamodel $\text{M}_i$ will have a $\xi(\text{M}_i)$ near zero and will also have $\pi(\text{M}_i)$ roughly equal to $\xi(\text{M}_i)$.
Each macroreplicate uses the same icon shape for $\xi(\text{M}_i)$ and $\pi(\text{M}_i)$ so they can be compared to each other and across the different software packages.

The range of the plots are selected to allow the reader to see the relationships between the $\xi$ and $\pi$ values across all packages. Several of the problematic values appear to the left of the plots, indicating that their values are too small to fit on the plot.
Values that are too large are shown to the right of the plots. All of the data values in the examples in this paper, including those that appear outside of the plot ranges, are available in the data in brief article associated with this paper \citep{EricksonDataInBrief}.


\subsection{The Borehole Function}

The borehole function described by \citet{Worley:Borehole} is commonly used for testing emulators \citep{morris1993bayesian}.
The input is 8-dimensional and each variable is confined to specified ranges.
The borehole function, $f(x)$, is
$$ f(x) = \frac{2\pi T_u (H_u - H_l)}{\log(r/r_w) [1+ \frac{2L T_u}{\log(r/r_w)r_w^2 K_w} + T_u/T_l]} . $$


\noindent We used an R implementation based on the one provided by \citet{simulationlib}, where they recommend
selecting sample points following a normal distribution for $r_w$,
a lognormal distribution for $r$,
and uniform distributions for the other six variables in their respective ranges.
We followed these recommendations for choosing sample points in each dimension, transforming them to be uniform in $[0,1]$.

\begin{figure}[!tbp]
  \centering
  \subfloat[$d=4$, $n=40$]{\includegraphics[width=0.5\textwidth]{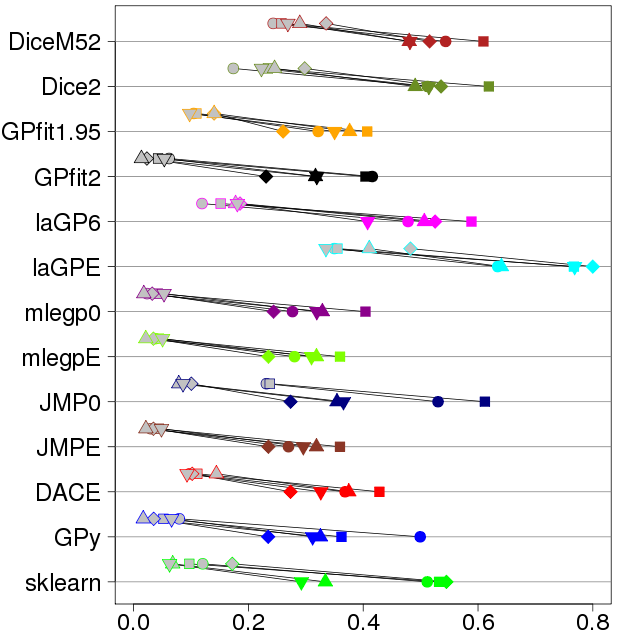}\label{Borehole1357_small}}
  \hfill
  \subfloat[$d=4$, $n=80$]{\includegraphics[width=0.5\textwidth]{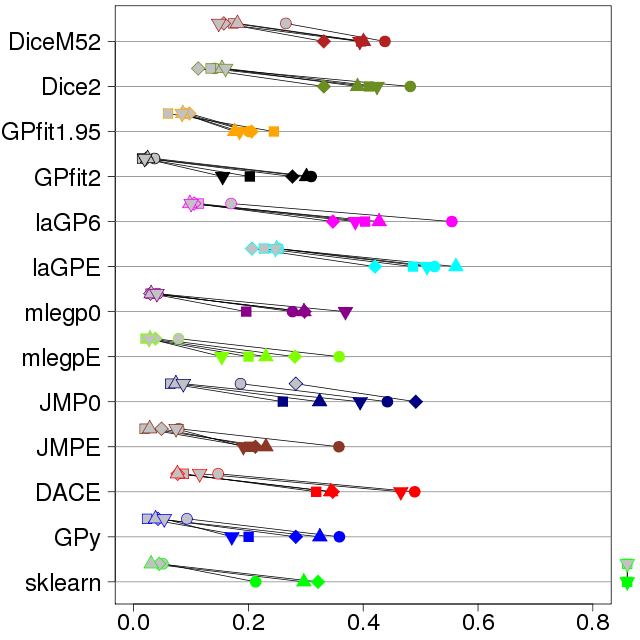}\label{Borehole1357_big}}
  \vfill
  \subfloat[$d=8$, $n=80$]{\includegraphics[width=0.5\textwidth]{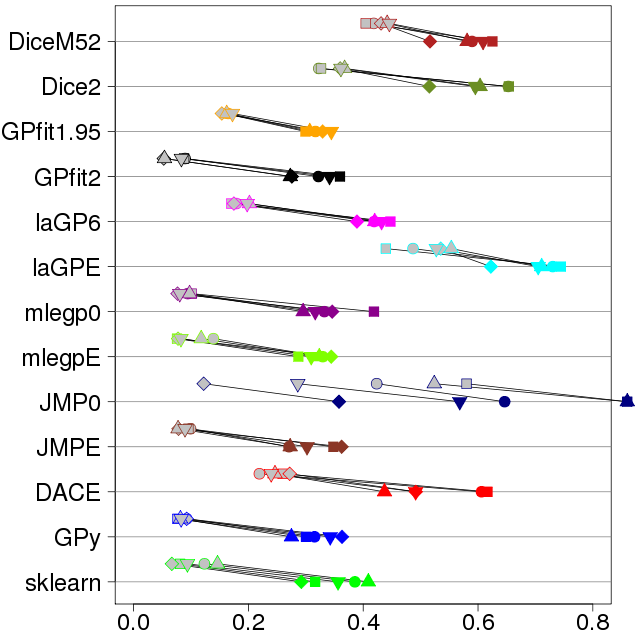}\label{Borehole_small}}
  \hfill
  \subfloat[$d=8$, $n= 160$]{\includegraphics[width=0.5\textwidth]{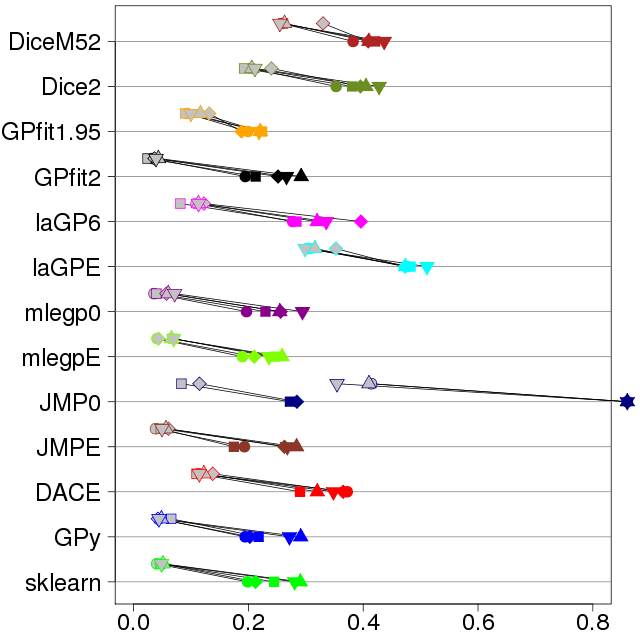}\label{Borehole_big}}
  \caption{Borehole 4-D and 8-D comparison. All four plots are on the same scale.}
  \label{Boreholeplots}
\end{figure}

We ran the full 8-dimensional function, and then projected that surface on the 4-dimensional subspace of
$r_w$, $T_u$, $T_l$, and $L$,
with the other values set to the middle value of their range.
Figure \ref{Boreholeplots} shows the results in plots of our comparisons.
The top row has the 4-dimensional function, the bottom row has the full 8-dimensional function, and we use two different sample sizes for each dimension.
The left column of plots in Figure \ref{Boreholeplots} has results for the smaller sample sizes ($n=40$ for 4 dimensions and $n=80$ for 8 dimensions).  The right column has results for the larger sample sizes ($n=80$ for 4 dimensions and $n=160$ for 8 dimensions).

When  the input sample size is increased by a factor of 2, the $\xi$ and $\pi$ values are typically reduced by about 30\%.
All four plots in Figure \ref{Boreholeplots} have been put on the same scale for easier comparison of this effect.
The GP metamodels clearly fit the borehole surface better than the linear model since almost all of the $\xi$ values are less than one. The exceptions are one macroreplicate of sklearn in Figure \ref{Borehole1357_big} and some of the JMP0 macroreplicates in Figures \ref{Borehole_small} and \ref{Borehole_big}; these have been cut from the plot and placed to the right to indicate that they could not fit on the plot without skewing the axes.

We can see that there is a problem in the error estimates for all of the packages.
For almost every macroreplicate, $\pi$ is less than $\xi$, often by a factor of two or more.
Users should be aware of the possibility of systematic underestimation of model error as seen in the borehole example.
These discrepancies between predicted errors and actual errors are likely due to the data not actually coming from a Gaussian process, so the surface does not match the model assumptions.
Methods such as cross-validation can be used to check for this problem.

Overall, GPfit, mlegp, JMPE, and GPy have the best performances on all four examples shown.
sklearn has trouble on some of the macroreplicates in four dimensions, but does better in eight dimensions.
DiceKriging, laGP, and DACE generally perform a little worse than the others, while JMP0 has some serious problems on the 8-dimensional surfaces.

\begin{figure}[!tbp]

  \subfloat[$n=80$]{\includegraphics[width=0.5\textwidth]{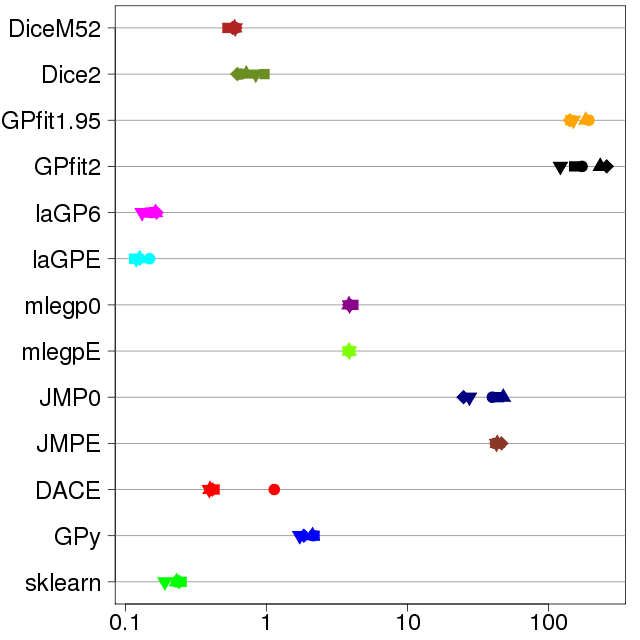}\label{Borehole02_RunTimes1}}
  \hfill
  \subfloat[$n=160$]{\includegraphics[width=0.5\textwidth]{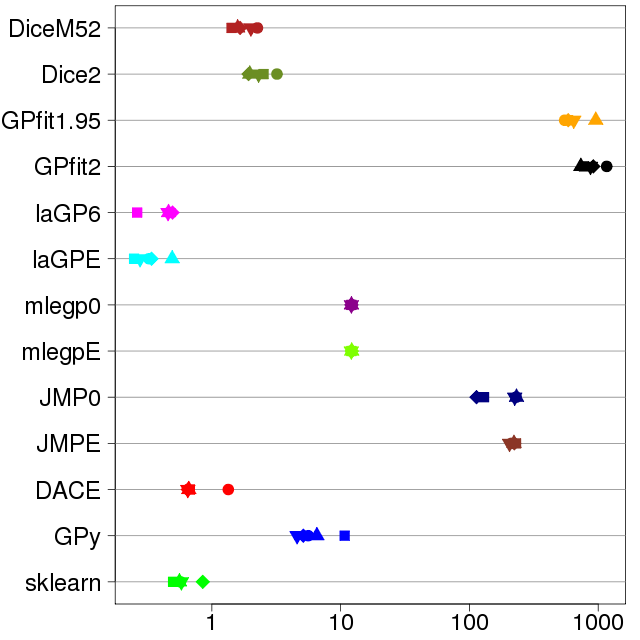}\label{Borehole02_RunTimes2}}
  \caption{Run times (seconds) for Borehole 8-D with $n=80$ and $n=160$, both with $m=2000$. There are enormous differences among the packages, but the relative speeds of the packages are similar.}
  \label{Borehole02_RunTimes}
\end{figure}

Figure \ref{Borehole02_RunTimes} shows how long (in seconds on a log scale) it took to fit each macroreplicate and make $m=2000$ predictions for the 8-dimensional borehole surface with $n=80$ and $n=160$ design points.
All macroreplicates for all packages were run on the same node of a Linux cluster, except for JMP which was run on a personal Dell laptop running Windows.
The relative run times for each package are the same for both sample sizes, and the same pattern is found on other test functions as well.
GPfit is by far the slowest, taking over 15 minutes per macroreplicate for $n=160$.
JMP is the next slowest, taking two minutes per macroreplicate, but this data is unreliable since it was run on a different computer.
The next slowest is  mlegp, taking about eight minutes, with GPy only slightly faster. The fastest packages were DiceKriging, laGP, sklearn, and DACE, which only took a handful of seconds. Thus we see that there is a massive difference in the run times, with a factor of over 1000 between the fastest and the slowest packages performing the same task.
The times shown in this plot are for the borehole function, but the relative times are similar for the other functions. In particular, GPfit and JMP are extremely slow, mlegp is also very slow, and the rest are much faster.
Therefore when one is choosing a package, it may be necessary to consider not only the model options and capability, but also how quickly it runs.
Run times must also be considered in the context of the data being used. If the data comes from a simulation model that takes hours per observation, then the difference of a minute may be negligible.

\subsection{The OTL Circuit Function}
\label{sectionotlcircuit}

\begin{figure}[!tbp]
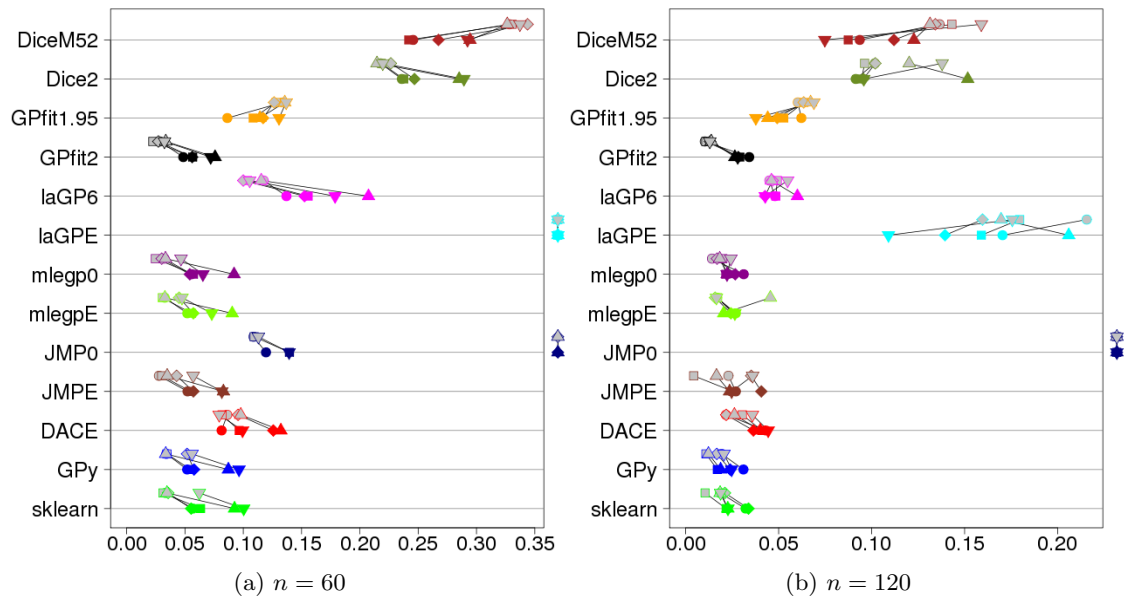

  \centering
  \subfloat[$n=60$]{\includegraphics[width=0.5\textwidth]{{{OTLCircuit2_SS60_RMSE_on_PRMSE_over_LM_stripchart_outlier_FINAL}}}\label{OTLCsmall}}
  \hfill
  \subfloat[$n=120$]{\includegraphics[width=0.5\textwidth]{{{OTLCircuit2_SS120_RMSE_on_PRMSE_over_LM_stripchart_outlier_FINAL}}}\label{OTLCbig}}
  \caption{OTL Circuit comparison}
  \label{OTLCplots}
\end{figure}

\citet{ben2007modeling} use a test function that describes an output transformer-less (OTL) push-pull circuit. There are six input parameters, five for resistors ($R_{b1}$, $R_{b2}$, $R_f$, $R_{c1}$, $R_{c2}$) and one for circuit gain ($\beta$).
The equation is given by
$$ V_m = \frac{(R_{b1}+0.74)\beta(R_{c2}+9)}{\beta(R_{c2}+9)+R_f} + \frac{11.35R_f}{\beta(R_{c2}+9)+R_f} + \frac{0.74R_f\beta(R_{c2}+9)}{(\beta(R_{c2}+9)+R_f)R_{c1}} $$

with $$ V_{b1} = \frac{12R_{b2}}{R_{b1}+R_{b2}} . $$
We used an R implementation provided by \citet{simulationlib}.
Figure \ref{OTLCplots} shows our results, based on five macroreplicates of $n=60$ and $n=120$ observations.

On this function, most of the fits  are much better than the linear model since most of the $\xi$ values are below 0.1.
Some of the laGPE and JMP0 points are placed to the right of the plot to indicate that their values are off the scale.
The problems exhibited by some of these packages, such as DiceKriging and laGPE, may be reduced by tuning the optimization parameters, but we did not attempt this as not all practitioners may  have this insight.
For both sample sizes, the best $\xi$ values come from GPfit2, mlegp, JMPE, GPy, and sklearn, with GPfit1.95, laGP6, and DACE performing only slightly worse.
DiceKriging, laGPE, and JMP0 perform poorly compared to the best packages.

The prediction errors, $\pi$ values, are fairly accurate on this data. Most of the $\pi$ values are less than the corresponding $\xi$ values by a small margin, but not by as much as in the borehole results of Figure \ref{Boreholeplots}.
Doubling the sample size reduced the $\xi$ and $\pi$ by about a factor of two, showing that increasing the sample size beyond $10d$ can have a large impact.

\subsection{Dette and Pepelyshev}
\label{sectiondetpep}

\citet{dette2012generalized} present an 8-dimensional model ``which is highly curved in some variables and has less curvature in another variables." The input is in $[0,1]^8$, and the output is given by the equation below:

$$ \eta(x) = 4(x_1-2+8x_2-8x_2^2)^2 + (3-4x_2)^2 +16\sqrt{(x_3+1)}(2x_3-1)^2+\sum_{k=4}^8 k\log{(1+\sum_{i=3}^kx_i)} .$$

\noindent This function and its R implementation were also taken from \citet{simulationlib}.

\begin{figure}[!tbp]
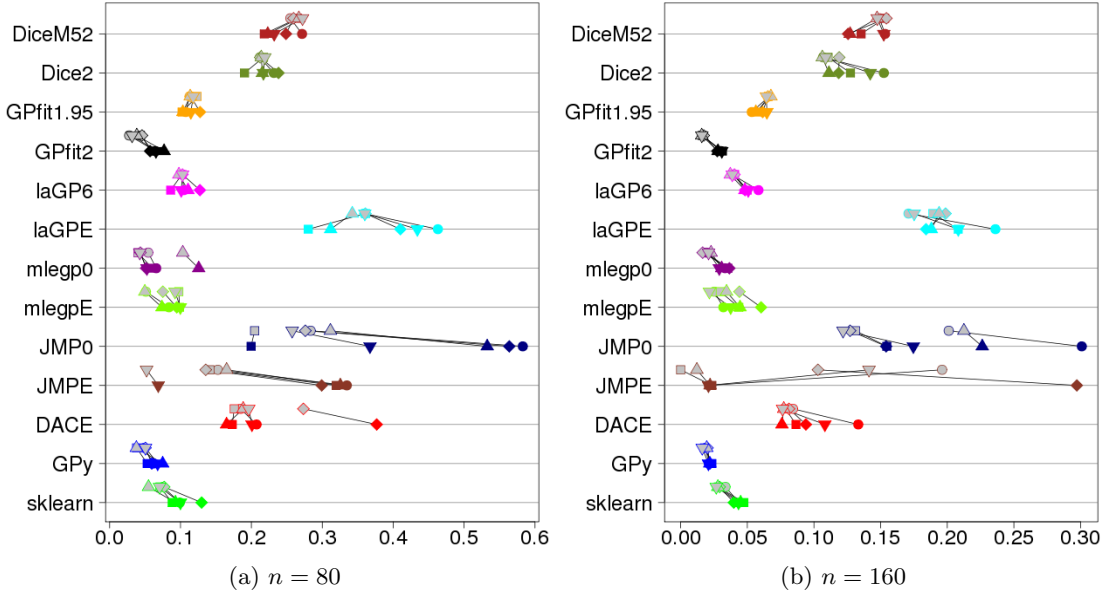

  \centering
  \subfloat[$n=80$]{\includegraphics[width=0.5\textwidth]{{{Detpep108d2_SS80_RMSE_on_PRMSE_over_LM_stripchart_outlier_FINAL}}}\label{Detpepsmall}}
  \hfill
  \subfloat[$n=160$]{\includegraphics[width=0.5\textwidth]{{{Detpep108d2_SS160_RMSE_on_PRMSE_over_LM_stripchart_outlier_FINAL}}}\label{Detpepbig}}
  \caption{Dette-Pepelyshev comparison}
  \label{Detpepplots}
\end{figure}

Figure \ref{Detpepplots} shows the results when we test this function with $n=80$ and $n=160$ design points in 8 dimensions. There are clear differences between the packages in these plots. GPy has the smallest $\xi$ values for both plots, with GPfit2 not far behind.
There is a large difference in the $\xi$ values between GPy and the worst performers, so again we see that the software used  makes a difference.
JMP0 is consistently bad on these examples, while JMPE is very inconsistent, with a mixture of good and bad fits.
laGP6 is generally very good, while laGPE, despite being consistent, is one of the worst performers.
GPfit, mlegp, GPy, and sklearn perform the best on this function.
The error predictions  for all packages except JMP are generally good, being within 25\% of the actual error.
We can also see that the performance ordering of the packages on this problem is similar to those for the OTL circuit example in Figure \ref{OTLCplots}.
Again increasing the number of observations beyond $10d$ had a decidedly beneficial effect, roughly halving the $\xi$ and $\pi$ values.

\subsection{Morris function}

The Morris function is a 20-dimensional function created by \citet{morris1991factorial},  and we use the version presented by
\citet{le2016metamodel}: 

$$
f(\bm{x}) = 
\sum_{i=1}^{20} \beta_i w_i +
\sum_{i<j}^{20} \beta_{i,j} w_i w_j +
\sum_{i<j<l}^{20} \beta_{i,j,l} w_i w_j w_l +
5 w_1 w_2 w_3 w_4.
$$

\begin{figure}[!tbp]
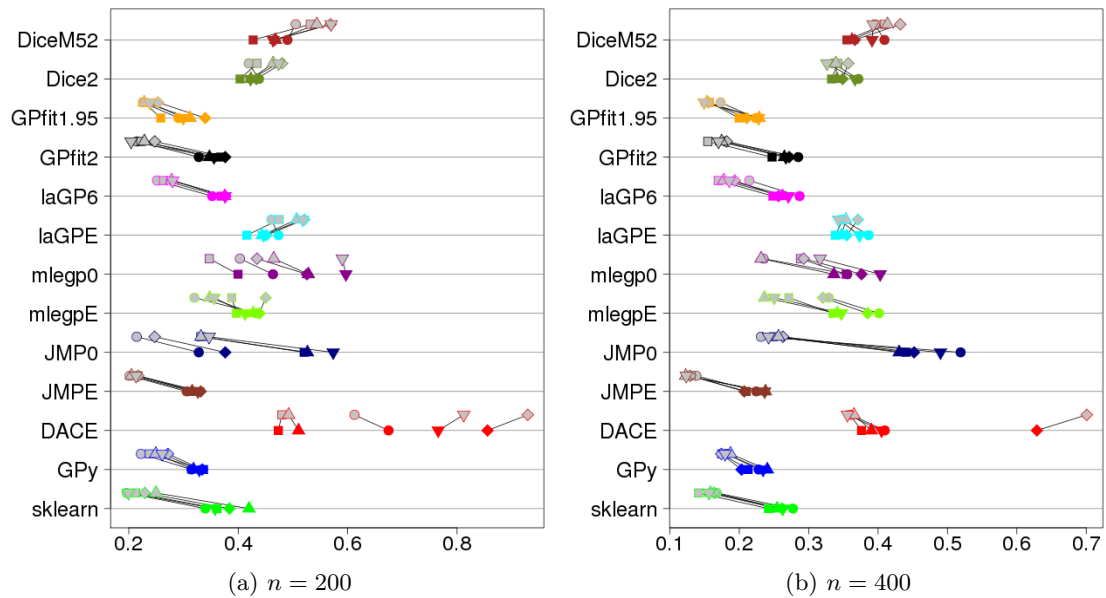

  \centering
  \subfloat[$n=200$]{\includegraphics[width=0.5\textwidth]{{{Morris1_SS200_RMSE_on_PRMSE_over_LM_stripchart_outlier_FINAL}}}\label{Morrissmall}}
  \hfill
  \subfloat[$n=400$]{\includegraphics[width=0.5\textwidth]{{{Morris1_SS400_RMSE_on_PRMSE_over_LM_stripchart_outlier_FINAL}}}\label{Morrisbig}}
  \caption{Morris comparison}
  \label{Morrisplots}
\end{figure}

\begin{sloppypar}
\noindent
Here, $\bm{x} \in [0,1]$,
${w_i = 2(1.1 x_i/(x_i+0.1) - 1/2)}$ for  $i=3,5,7$, and
$w_i = 2(x_i - 1/2)$ for all other values of $i$.
The coefficients are $\beta_i = 20$ for $i=1,..., 10$, $\beta_{i,j}=-15$ for $i,j=1,...,6$, $\beta_{i,j,l}=10$ for $i,j,l=1,..., 5$.
All other coefficients are set  to $\beta_i = (-1)^i$, $\beta_{i,j} = (-1)^{i+j}$
and $\beta_{i,j,l}=0$.
The results from using the Morris function are shown in Figure \ref{Morrisplots} using input samples of size $n=200$ and $n=400$.
\end{sloppypar}



DACE has fitting problems, especially for $n=200$, but the rest are fairly consistent.
The $\xi$ values are all fairly large, many around 0.4.
This demonstrates that in higher dimensions it is more difficult to get a fit that is substantially better than the linear model, especially when the function is relatively linear.
GPfit, laGP6, JMPE, GPy, and sklearn perform the best, but they all underestimate the error by a significant amount.

\section{Recent literature / Advanced models}

In this paper, we have focused on an ordinary GP model. However, there are many variations of this model that can be used in situations where there is domain knowledge about the data or where the basic model is inadequate. If the data is noisy then a nugget should always be used and estimated. There are also many different correlation functions beyond the Gaussian and Matérn that may perform better with certain types of data. If the data set is large then there are approximation models that should be used instead, since they will run much faster with a small loss of accuracy. If there are $n$ design points, then the computation complexity for kriging is  $O(n^3)$, which is far too slow for modern problems with millions of data points.

\subsection{{Stochastic Kriging}}
\label{sectionstoch}

While computer experiments often assume that there is no variability in the data, this is not the case in stochastic simulations.
When the noise is similar across the entire response surface, then the basic model should suffice by using the nugget term.
However, when the noise level varies across the surface, called heteroscedasticity, stochastic kriging should be used.
Stochastic kriging, by \citet{ankenman2010stochastic}, accommodates for noise in data collected by assuming that the variance in the data is different at each design point.
In order to estimate the noise at each point, replicates must be collected at every point in the design.
This is equivalent to having a different nugget at each design point.
Thus instead of adding $\delta I$ to the diagonal of the correlation matrix,
$\text{diag}(\bm{\delta})$ is added, where $\delta_i \propto \text{Var}(x_i)$.
This requires  $\text{Var} (x_i)$ to be estimated by replicates at each unique design point.

Stochastic kriging has been used for modeling simulation data from many fields, such as in game theory simulations \citep{pousi2010game} and finance for measuring portfolio risk \citep{liu2010stochastic}.
Stochastic kriging models are often run in two stages.
In the first stage, a small number of samples are taken for all design points.
For the second stage, the number of samples for each point is allocated according to Equation (29) in \citet{ankenman2010stochastic}, which puts more replicates at points that have large sample variances and are centrally located.

Of the software packages discussed above, only mlegp and DiceKriging are able to perform stochastic kriging.
For each package a variance estimate at each point must be provided.
In mlegp, this vector is passed as the ``nugget'' parameter, and the diagonal of the nugget matrix is set to be proportional to these values \citep{dancik2011mlegp}.
The nugget scaling parameter is estimated along with the other parameters.
In DiceKriging, this same vector as passed as the ``noise.var'' parameter
\citep{roustant2012dicekriging}.
The prototype software developed for the paper \citet{ankenman2010stochastic} used off-the-shelf optimization algorithms that do not scale to larger problems and often have convergence issues. It will not be considered in this comparison.

To demonstrate the use of mlegp and DiceKriging for stochastic kriging, we use data taken from the standard M/M/1 queue model.

\subsubsection{M/M/1 Queue}

\begin{figure}[!tbp]
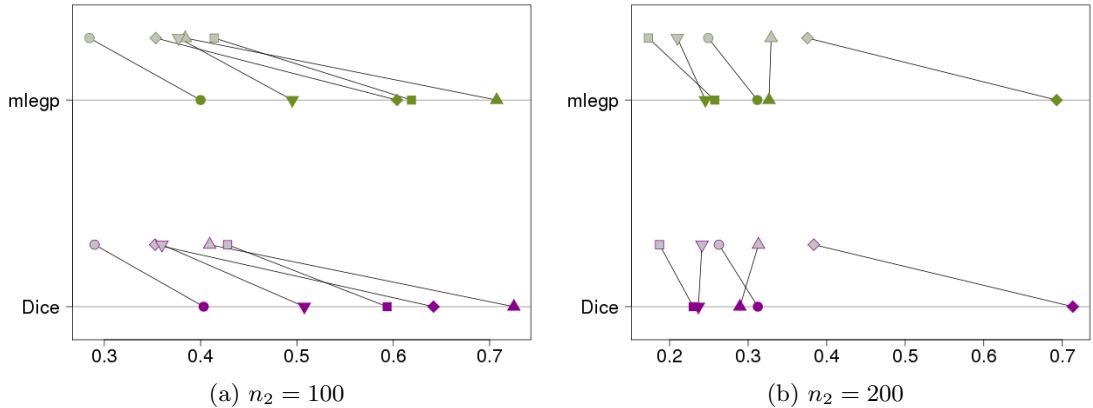

  \centering
  \subfloat[$n_2=100$]{\includegraphics[width=0.5\textwidth]{{{MM1_RMSE_on_PRMSE_stripchart100}}}\label{MM1L}}
  \hfill
  \subfloat[$n_2=200$]{\includegraphics[width=0.5\textwidth]{{{MM1_RMSE_on_PRMSE_stripchart200}}}\label{MM1R}}
  \caption{M/M/1 stochastic kriging comparison using mlegp and DiceKriging. Both use 5 samples at each point in the first stage, and then the total number of points allocated in the second stage is $n_2=100$ for Figure \ref{MM1L} and $n_2=200$ for Figure \ref{MM1R}.}
  \label{MM1plot}
\end{figure}

The M/M/1 queue is a service model that represents a system with one server and interarrival and service times that are independently exponentially distributed. We set the service rate $\lambda=1$ and the arrival rate $0 \leq x < 1$.
We model the number of customers waiting in the queue as a function of the arrival rate, which is known to have mean $y(x)=x/(1-x)$ and variance $x/(1-x)^2$.
For design points we use seven equally-spaced points at (0.3, 0.4, 0.5, 0.6, 0.7, 0.8, 0.9).
In the first stage, we take $n_1=5$ samples at each point. Then the second stage is run with a total of $n_2=100$ and $n_2=200$ points allotted to the individual design points according to the square root of the variance of the output for the first stage samples.
The results are shown in Figure \ref{MM1plot},
where there is no distinguishable difference between the two packages.
Even with $n_2=100$ observations the $\xi$ values are relatively large, showing that stochastic kriging typically needs more observations to fit a surface.
Both packages tend to underestimate the error when the sample size is small.

\subsection{Other models}

Many advances in Gaussian process fitting have been adapting the method to be suitable for large amounts of data, such as by exploiting sparsity.
\citet{snelson2005sparse} present a sparse GP method that reduces the size of the covariance matrix by using pseudo-input points.
\citet{HensmanFL13} produce a method that works for data sets with millions of data points, which would be prohibitively slow for the standard model.
Sparsity can be induced into the correlation matrix by having a correlation function with compact support, meaning that it is zero for points that are sufficiently far away \citep[][p. 87-8]{GPMLBook}. These functions usually are piecewise polynomials that resemble the Gaussian correlation function.
\citet{BG:LGPA} provides a way to fit GPs to large experiments quickly. Their model induces sparsity by only selecting the points that provide the greatest reduction in predictive variance when calculating the metamodel function at a given point. They also allow for quick sequential updating and trivial parallelization, making their method very practical. \citet{laGP:cran} has provided the R package laGP that implements most of these methods, in addition to the basic model that we explore in this paper.
Further work in \citet{sunggramacyhaaland} makes the search for the best sub-design much faster.

The recent paper by \citet{binois2016practical} provides a significant improvement to stochastic kriging.
They provide a more favorable framework by putting the problem in an inferential scheme with a single objective and explicit derivatives.
They introduce some smoothing techniques that allow
design points without replicates to be used in the model, which is a shortcoming of previous versions of stochastic kriging.
Also, they use the Woodbury identity \citep{woodbury1950inverting} to ensure that the computational complexity is similar to other stochastic kriging methods.

Advances have also come by combining the Gaussian process with other models. Neural networks inspired the work of \citet{Damianou:deepGPs13}, who present a deep Gaussian process model that uses hierarchical Gaussian process mappings. \citet{gramacy2012bayesian} add GPs to the Bayesian partition tree model of \citet{chipman1998bayesian} so that a GP is fit to each partition.
\citet{williams1998bayesian} create a model that uses GPs for Bayesian classification.
There have also been advances by allowing the model to take categorical input.
\citet{platt2001learning} use GPs with categorical input to generate music playlists.
\citet{chen2013stochastic} address the use of stochastic kriging with categorical input.

Some of the packages we investigate have advanced models available for users. GPy has many models, including classification, sparse regression, latent variable models, and more.
In addition to providing GP classification models, scikit-learn also has other machine learning models such as clustering, neural networks, and support vector machines.
laGP provides the approximate GP model, as explained above, that is useful for massive data sets.
When choosing a software package, users should consider the depth of options available on the platform, and what types of models they would potentially use beyond the basic GP model to get better results.

\section{Discussion}

\subsection{Summary of packages}
\label{sumofpackages}
In this paper, we study various Gaussian process fitting software packages, see Table \ref{packagetable}, and compare their performance using the GP model with Gaussian correlation for global fitting.
We do not compare their performance for optimization, i.e., accurately locating an optimal point.
We assess them based on the quality of their response predictions and their error estimates, each of which are averaged across the region of interest.
Other possible criteria that we do not consider include maximum error and relative error.
In many cases the different packages give similar, or even indistinguishable, results---which is expected since they are using the same models on the same surfaces.
However, due to differences in the parameter estimation routines, the packages often give different results on complicated surfaces.


\textbf{DiceKriging}, an R package, performed somewhat worse than many of the other packages in our examples.
However, DiceKriging runs faster than the R packages mlegp and GPfit, and provides more customization options than laGP.
It has multiple correlation functions available and can estimate a nugget.
By default the correlation function is the Mat\'{e}rn with $\nu=5/2$, but we did not see a large difference between that and the Gaussian correlation function.
DiceKriging also provides functionality for stochastic kriging, as demonstrated in Section \ref{sectionstoch}.

Another R package, \textbf{GPfit}, uses the most extensive parameter optimization, which makes it very dependable. In our tests we found GPfit to be reliable and give good results. The cost of this is that it takes significantly longer to use on larger data sets, taking noticeably longer on samples larger than even 50.
On the borehole test, each macroreplicate for GPfit for a sample size of 500 took over two hours, while all the other packages finished in minutes or even seconds.
For this reason we do not recommend using GPfit when the data set is large and time is valuable.
We ran GPfit with its default exponent of 1.95 in the correlation function and also used the Gaussian process model where that exponent is 2.00. In general we did not see a large difference between the two in performance or run time. GPfit uses 1.95 as the default because it is supposed to provide computational stability.

The R package \textbf{laGP} is the fastest package we tested. We used it with estimating a nugget and with setting a small nugget, the latter of which gave better results.
The main benefit of laGP is that it runs very quickly, especially when repeatedly adding data in a sequential manner.
In addition, it provides some additional complex models that are useful for large data sets. Thus we do not recommend laGP for kriging with small sample sizes, but we do suggest looking into its advanced functionality if there are thousands of sample points.

The final R package we evaluated,
\textbf{mlegp}, performed well in our testing.
We used it with both setting the nugget to zero and estimating the nugget, and did not see a large difference.
In addition, mlegp was a little slower than the other packages, though not as slow as GPfit.
One benefit of mlegp is that it also can do stochastic kriging, as shown in Section \ref{sectionstoch}.

\textbf{JMP} is a commercial software platform that makes data analysis easy for practitioners.
When the nugget was set to zero, JMP performed poorly on most of our test functions.
When the nugget was estimated, however, JMP performed substantially better, often on par with the best packages.
JMP seems to run relatively slow.
For our testing, JMP was run on a laptop and the other packages were run on a cluster. However in our experience, the other packages (except for GPfit) ran faster when also run on the same laptop.
Thus users should be careful when using JMP, particularly when the nugget is not estimated, since the results may be spurious, and users might get better and faster results using a different software option.

The Matlab toolbox \textbf{DACE} was fast but generally provided a slightly worse fit than the best models.
DACE is very basic and has not been updated in years, so we recommend using other packages if there is a desire to progress to more advanced models.

\textbf{GPy}, a Python module, gave the best fitting results in most of our tests.
It was an order of magnitude slower than the fastest package, and was generally in the middle in terms of speed. GPy also provides many options and advanced models that the practitioner can explore once they have mastered the basic model.

Finally, the Python module \textbf{scikit-learn} contains GP fitting capability in addition to many other machine learning algorithms.
It was near the best on most examples, but occasionally exhibited inconsistency.
In preliminary tests, we found that it performed better when the data is scaled and more optimization restarts are used.
It was also one of the fastest packages.
Although we only included the results using scikit-learn with the Gaussian correlation function in this paper,
we have found in some of our tests that using the Mat\'{e}rn correlation function gives better results.
Although scikit-learn does not provide advanced GP models, it does provide other machine learning models such as support vector machines and random forests. Thus it would be a useful tool for those who want to use a single platform for multiple machine learning methods.

\subsection{Conclusion}
This paper focuses on the traditional Gaussian process model with Gaussian correlation.
Despite specifying the same type of GP metamodel, we found that there are often significant differences between the metamodel predictions made by various software packages on the same data.
Practitioners should be aware of the quality of predictions,  typical run time, and model options when choosing a modeling software to use.

There are many modifications of the model and other correlation functions that will often give better and faster results if there is prior knowledge about the structure of the data or if there is a large number of observations. We focus on the simple model because it is used by many practitioners who do not want to spend the time to learn the intricacies of the model, but wish to use the power of GP fitting.  However, if unstable or nonsensical results are observed when fitting the simple GP model, we encourage modelers to consider using the packages with more advanced features that we allude to in Section \ref{sumofpackages}.


\section{Acknowledgments}

U.S. Department of Defense Distribution Statement: Approved for public release; distribution is unlimited. The views expressed in this document are those of the authors and do not reflect the official policy or position of the Department of Defense or the U.S. Government.  This work was supported in part by the Office of Naval Research via NPS’s CRUSER initiative, the NPS Naval Research Program NPS-17-N191-B, and the Naval Supply Systems Command Fleet Logistics grant number N00244-15-2-0004.

\bibliographystyle{elsarticle-num-names-alpha} 
\bibliography{MyBib}

\end{document}